\documentclass[prl,twocolumn,superscriptaddress,showpacs,debugon,nofootinbib]{revtex4}
\usepackage{epsfig}

\newcommand{\bk}{{\bf k}}
\newcommand{\be}{\begin{equation}}
\newcommand{\ee}{\end{equation}}
\renewcommand{\r}{{\bf r}}

\newcommand{\R}{{\bf R}}

\newcommand{\ep}{\epsilon}

\begin{document}


\title{Effect of superradiance on transport of diffusing photons in cold atomic gases}
\author{A. Gero and E. Akkermans}
\affiliation{Department of Physics, Technion Israel Institute of Technology,
  32000 Haifa, Israel}

\begin{abstract}
We show that in atomic gases  cooperative effects like superradiance and subradiance  lead to a  potential between two atoms that decays like $1/r$. In the case of superradiance, this potential is attractive for close enough atoms and can be interpreted as a coherent mesoscopic effect. The contribution of superradiant  pairs to multiple scattering properties of a dilute gas, such as photon elastic mean free path and group velocity, is significantly  different from that of independent atoms. We discuss the conditions under which these effects may be observed and compare our results to recent experiments on photon transport in cold atomic gases.
\end{abstract}


\date{\today}

\maketitle


The issue of coherent multiple scattering of photons in cold atomic gases is important since it presents a path towards the onset of Anderson
localization transition, a long standing and still open issue.  The large resonant scattering cross-section of photons reduces the elastic mean free path to values comparable to the photon
wavelength for which the weak disorder approximation breaks down thus signaling the onset of Anderson localization transition \cite{am,rama}. Another advantage
of cold atomic gases is that sources of decoherence
and inelastic scattering such as Doppler  broadening can be
neglected. Moreover,  propagation of photons in atomic gases differs  from the case of electrons in
disordered metals or of electromagnetic waves in
suspensions of classical scatterers for which mesoscopic effects and Anderson localization have been thoroughly investigated \cite{am}. This problem is then of great interest since it may raise new issues in the Anderson problem such as change of universality class and therefore new critical behavior.  New features displayed by the photon-atom problem  are the existence of  internal
degrees of freedom (Zeeman sublevels) and cooperative effects such as subradiance or superradiance that lead to effective interactions between atoms \cite{ketterle}. These two differences may lead to qualitative changes of  both mesoscopic quantities and Anderson localization. Some of the effects  of a  Zeeman degeneracy have  been
investigated in the weak
disorder limit \cite{miniat} using a set of
finite phase coherence times \cite{amm} which reduce mesoscopic effects, such as coherent backscattering \cite{am,akkmayn}.  The aim of this paper is to investigate the influence of cooperative effects and more specifically of superradiance on the multiple scattering of photons.  We show that two atoms in a Dicke superradiant state \cite{dicke} interact by means of a  potential which, once averaged over disorder configurations, is attractive at short distances and decays like $1/r$. This potential, analogous to the one considered in \cite{thiru, kurisky}, has important consequences  on transport properties since the contribution of superradiant pairs  of atoms in a dilute gas  provides smaller values of  both group velocity and  diffusion coefficient  so that the photons become closer to the edge of Anderson localization.

Atoms are taken as degenerate two-level systems denoted by $|g \rangle= |j_{g}=0,m_{g}=0
\rangle$ for the ground state and $|e \rangle = |j_{e}=1,m_{e} \rangle$ for  the excited state,
where $j$ is the total angular momentum and $m$ is its  projection
on the quantization axis, taken as the
$\hat{z}$ axis. The energy separation between the two levels including radiative shift  is
$\hbar\omega_{0}$ and the natural width of the excited level is $\hbar \Gamma$. We consider a pair of such atoms in an external radiation field and the corresponding 
Hamiltonian is $H = H_0 + V$, with \be
H_{0}={\hbar\omega_{0} \over 2} \sum_{l=1}^{2}(|e\rangle\langle
e|-|g\rangle\langle g|)_{l}+ \sum_{\bf{k}\ep} \hbar
\omega_{k} a_{\bf{k}\ep}^{\dag}a_{\bf{k}\ep} \label{eq1} \ee
$a_{\bf{k}\ep}$ (resp. $a_{\bf{k}\ep}^{\dag}$) is the annihilation (resp. 
creation) operator of a mode of the field of wave vector $\bf{k}$,
polarization $\hat \varepsilon_{\bf{k}}$ and angular frequency
$\omega_{k}=c|\bf{k}|$.  The interaction $V$ between the
radiation field and the dipole moments of the atoms may be
written as \be V=- \bf{d}_{1}\cdot\bf{E}(\bf{r_{1}}) -
\bf{d}_{2}\cdot\bf{E}(\bf{r_{2}}) \end{equation} where $\bf{d_{l}}$
is the electric dipole moment operator of the $l$-th atom and  ${\bf E}(\r
)$ is the electric field operator.

The absorption of a photon by a pair of atoms in their ground state, leads to a configuration where the two atoms, one excited and the second in its ground state, have multiple exchange of a photon,  giving rise to an effective interaction potential
and to a modified lifetime as compared to independent atoms. These two quantities are obtained from  the matrix elements
of the evolution operator $U(t)$ between states such as $|g_1 e_2 ;
0 \rangle$.  There are  six unperturbed and degenerate states with no photon, given by  $\{ |g_{1}  e_{2i} ; 0 \rangle, |e_{1j} g_{2} ; 0
\rangle \}$  in a standard basis where $i,j=-1,0,1$.  The symmetries of the Hamiltonian, 
namely its invariance by rotation around the axis between the two
atoms, and by reflection with respect to a plane containing
this axis, allows one to find
combinations of these  states that are given by $|\phi_i ^\epsilon \rangle =
{1 \over \sqrt{2}} [ |e_{1i} g_2 ; 0 \rangle + \epsilon  |g_{1}
e_{2i} ; 0 \rangle ] $ with $\epsilon = \pm 1$, so that $\langle \phi_j ^{\epsilon'} | U(t) |
\phi_i ^\epsilon \rangle = \delta_{ij} \delta_{\epsilon \epsilon'}
S_i ^\epsilon (t) $ and  \be S_i ^\epsilon (t) = \langle e_{1i} g_2
; 0 | U(t) | e_{1i} g_2 ; 0 \rangle + \epsilon \langle g_{1} e_{2i}
; 0 | U(t) | e_{1i} g_2 ; 0 \rangle \label{eq3} \ee The states
$|\phi_i ^\epsilon \rangle$ are the well-known Dicke states, otherwise defined as $|L M
\rangle$, where $L$ is the cooperation number and $M$ is half of 
the total atomic inversion \cite{dicke}  so that $ |\phi_i
^+ \rangle = |1 0 \rangle $ and $|\phi_i ^- \rangle = |0 0 \rangle$.
For large times, $t \gg r/c$, where $r$ is the
distance between the two atoms,  up to second order in the
coupling to the radiation, we obtain that 
\be S_i ^\epsilon (t) \simeq 1 - {it
\over \hbar} [ \Delta E_i ^\epsilon - i {\hbar \Gamma_i ^\epsilon \over 2}] .
\label{eq4} \ee 
The two real quantities $\Delta E_i ^\epsilon $ and
$\Gamma_i ^\epsilon $ are respectively the 
interacting potential  and the probability per
unit time of emission of a photon by  the two  atoms in
a Dicke state $|\phi_i ^\epsilon \rangle$.  A standard calculation \cite{stephen} gives 
\be \Delta E_i ^\ep  = \ep {3
\hbar \Gamma \over 4} \left[ - p_i {\cos k_0 r \over k_0 r} + q_i \left( {
\cos k_0 r \over (k_0 r)^3 } + {\sin k_0 r \over (k_0 r)^2} \right)
\right] \label{eq5} \ee and \be {\Gamma_i ^\epsilon \over  \Gamma} = 1 - {3
\over 2} \ep \left[ - p_i {\sin k_0 r \over k_0 r} + q_i \left(
{\sin k_0 r \over (k_0 r)^3} -  { \cos k_0 r \over (k_0 r)^2 }
\right) \right] \label{eq6} \ee where $k_0 = \omega_0 / c$. We have defined 
$
p_i = 1 - { \bf \hat r}^2 _i $ and $ q_i = 1 -  3 {\bf \hat r}^2 _i 
$, $\bf \hat r$ being a unit vector along the two atoms.  At short distance $k_0 r \ll 1$, we obtain that 
$\Gamma_i ^+ = 2 \Gamma$ for the superradiant state $|\phi_i
^+ \rangle = |1 0 \rangle$ and $\Gamma_i ^- = 0 $ for the subradiant state $|\phi_i ^- \rangle = |0 0 \rangle$.

For a photon of wavevector $\bk$ incident on an atomic cloud, the potential we shall denote by $V_e$ 
is obtained by averaging upon the random orientations of the pairs
of atoms.  
 Since $\langle q_i \rangle =0$ and
$\langle p_i \rangle = 2/3$ regardless of $i$, we obtain for  the average potential $V_{e}$ \be \ep
V_{e} (r) = \langle \Delta E_i ^\ep \rangle  = - \ep {\hbar \Gamma \over
2} {\cos k_0 r \over k_0 r} \label{eq7} \ee and the average inverse
lifetimes of  Dicke states are \be\langle \Gamma_i
^\epsilon  \rangle= \Gamma \left[ 1 + \ep {\sin k_0 r \over k_0 r}
\right] \label{eq8} \ee which retains the same features as (\ref{eq6}) for $k_0 r \ll 1$. 

Let us characterize the interaction potential $V_e$.  Whereas for a single pair of atoms,  the potential (\ref{eq5})  is anisotropic and
decays at short distance like $1/ r^3$, a behavior that originates from 
the transverse part of the photon propagator, we obtain that on average over angular configurations, the potential (\ref{eq7}) between two atoms in a Dicke state $M=0$ becomes  isotropic and decays 
like $1/ r$. This behavior is also obtained by considering the interaction of two-level atoms with a scalar wave.  This  could have been anticipated since the transverse contribution $q_i$ to the photon propagator averages to 0.  A similar  expression for the interacting potential has been obtained for  the case of  an intense radiation
field \cite{thiru, kurisky}.
 But this latter  potential is fourth order in the coupling to the radiation and it corresponds to the interaction energy between two atoms in their ground state in the presence of at least one photon. The average potential $V_e$ we have obtained is different. 
It is  second order  in
the coupling to the radiation and it corresponds to the interaction energy of Dicke states $M=0$ in vacuum.  


We turn now to scattering properties of  Dicke states. The collision operator is given by $T (z) = V + V G(z) V$ where $V$
is given by (2) and $G(z)$ is the resolvent  whose expectation value in the Dicke state $M=0$ is obtained by a summation of the series of exchange of a virtual photon
between the two atoms. The matrix element that describes the
transition from the initial state $|i \rangle = |1-1;
{\bk} {\hat \varepsilon} \rangle$ where the two atoms  are in their
ground state in the presence of a photon $({\bk} {\hat \varepsilon})$ to the
final state $|f \rangle = | 1-1; {\bk}^{'}{\hat \varepsilon}^{'} \rangle$ is the sum of the
superradiant and subradiant contributions, $ T = T^{+} + T^{-}$, with $T^{\pm} = \langle f | V | \phi^{\pm} \rangle
\langle \phi^{\pm}| G ( \omega - \omega_0) | \phi^{\pm}
\rangle \langle \phi^{\pm} | V | i \rangle$ \cite{rk}. A standard
derivation leads to the following expressions for the average amplitudes $T_{e}^{\pm}$
    \be T_{e} ^{+} = A e^{ i
(\bk - \bk') \cdot \R} \cos \left( {\bk \cdot \r \over 2} 
\right)\cos \left( {\bk' \cdot \r \over 2}  \right) G_{e} ^{+} \label{eq11}
\ee and  \be T_{e} ^{-} = A e^{ i (\bk - \bk') \cdot \R } \sin
\left( {\bk \cdot \r \over 2} \right)\sin \left( {\bk' \cdot \r \over 2}  \right) G_{e} ^{-} .\label{eq12} \ee We have defined  $\r =
\r_1 - \r_2$ , ${\bf R} = ({\bf r_1} + {\bf r_2}) / 2$ and $A =  {\hbar
\omega \over \ep_0 \Omega} d ^2 ({\hat e}_j \cdot {\hat \varepsilon} )
({\hat e}^* _j \cdot {\hat \varepsilon}'^* ) $ ($d$ is a reduced
matrix element and $\Omega$ the quantization volume). The average propagators $G_{e} ^{\pm}$ 
associated respectively to the superradiant and subradiant states are, \be
G_{e} ^{\pm} = \langle \phi^{\pm}| G( \delta) | \phi^{\pm}
\rangle = {1 \over \hbar (\delta + i {\Gamma \over 2} \pm {\Gamma
\over 2} {e^{i k_0 r} \over k_0 r}) } \label{eq13} \ee where close to resonance, $\delta
= \omega-\omega_{0} \ll \omega_{0}$ and where we have used (\ref{eq7}) and (\ref{eq8}) for the average potential and for the average inverse lifetimes. At short distances $k_0 r \ll 1$,
the subradiant amplitude $T_{e} ^{-}$ becomes
negligible as compared to the superradiant term (\ref{eq11}). Therefore, the potential (\ref{eq7}) is attractive and decays like $1/r$. We can interpret these results
by saying that, at short distances $(k_0 r \ll 1)$,  the time evolution of the initial
state $| \psi (0) \rangle = |e_1, g_2; 0 \rangle = {1 \over
\sqrt{2}} [ |\phi^{+} \rangle + |\phi^{-} \rangle ]$ corresponds for
times shorter than $1 / \Gamma$  to a periodic exchange of a
virtual photon between the two atoms at the Rabi frequency 
$ (\langle \Delta E^- \rangle -  \langle \Delta E^+  \rangle) / \hbar \simeq \Gamma / (k_0 r)$ which
is much larger than $ \Gamma $. For larger times,  the two
atoms return to their ground state and a real photon $(\bk' {\hat
\varepsilon}')$ is emitted. At large distances $(k_0 r \gg 1)$, the
Rabi frequency becomes smaller than $ \Gamma$, so that the
excitation energy makes only a few oscillations between the two
atoms, thus leading to a negligible interaction potential \cite{ga}. 



It is interesting to derive the previous results in another way that emphasizes the analogy with weak localization corrections \cite{am,rama}. 
\begin{figure}[ht]
\centerline{ \epsfxsize 4cm \epsffile{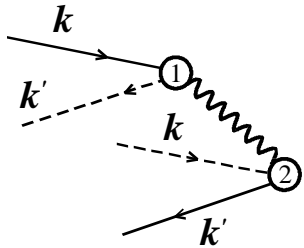} } \caption{\em Diagrammatic representation of the two amplitudes that describe double scattering of a scalar wave. 
The wavy line accounts for the photon exchange between the two atoms. This diagram is known in quantum mesoscopic physics as a  Cooperon.
}
\label{fig1}
\end{figure}

To that purpose,
we write the scattering amplitude $T$ defined previously as a superposition of  two 
scalar amplitudes $T_1$ and
$T_2$ \cite{rk2},  each of them being  a
sum of single scattering and double scattering contributions, that is  \be T_1 = {t \over 1 - t^2 G_0 ^2} \left[ e^{i
(\bk - \bk') \cdot \r_1} + t G_0 e^{i (\bk \cdot \r_1 - \bk' \cdot
\r_2)} \right] \label{eq14} \ee and \be T_2 = {t \over 1 - t^2 G_0
^2} \left[ e^{i (\bk - \bk') \cdot \r_2} + t G_0 e^{i (\bk \cdot
\r_2 - \bk' \cdot \r_1)} \right] . \label{eq15} \ee Here $t = ( 2 \pi \Gamma
/  k_0) / ( \delta + i \Gamma / 2)$ is the amplitude of a scalar wave scattered by a single
atom at the origin and the prefactor $t / ( 1 - t^2 G_0 ^2)$ where $G_0 = -e^{ i
k_0 r} / 4 \pi r$ accounts for the summation of the series of virtual 
photon exchange between the two scatterers.  We single out  in the total amplitude $T= T_1
+ T_2$  the single scattering contribution $T_{s}$ and write the intensity associated to the double
scattering term shown in Figure \ref{fig1} as \be |T  - T_{s}|^2  = 2 \left| {t^2
G_0 \over 1 - t^2 G_0 ^2 } \right|^2 \left[ 1 + \cos (\bk + \bk' )
\cdot (\r_1 - \r_2 ) \right] . \label{eq16} \ee We 
recognize in the bracket  the well-known Cooperon interference
term which is at the basis of coherent effects in quantum
mesoscopic systems such as weak localization and coherent backscattering \cite{am,rama,akkmayn}. The interference term reaches its
maximum value 1 for $\r_1 = \r_2$ 
so that we obtain from (\ref{eq14})
and (\ref{eq15}) that $T_1 = T_2 \propto (1/2) T_{e} ^+$, up to a proportionality  factor \cite{rk2}. Thus, the total
amplitude is exactly given by the superradiant term with no subradiant  contribution.

We consider now multiple scattering of a photon by  superradiant pairs built out of atoms separated by a distance $r$ and coupled by the attractive interaction potential $V_{e}$. This situation corresponds to a dilute gas that  fulfills $r \ll \lambda_0 \ll n_i ^{-1/3}$ where $n_i$ is the density of pairs and $\lambda_0 = 2 \pi / k_0$ is the atomic transition wavelength. Based on this inequality, we may consider the two atoms that  form  a superradiant pair through exchange of a virtual photon as an effective scatterer and neglect cooperative interactions between otherwise well-separated pairs.  The photon behavior  is described by the configuration average of its Green's function whose expression is obtained from a standard derivation \cite{am}.  In the limit of large enough densities of weakly scattering pairs, it reduces to the calculation of a 
self-energy given  in terms of the average propagator (\ref{eq13}) by  
\be 
\overline{\Sigma}_{e}
^{(1)} =  {6 \pi \hbar \Gamma n_i \over k_0} {\overline  G}_e ^+ =  { 6 \pi n_i  \over k_0 r_m} \int_0^{r_m} {dr  \over {
\delta \over \Gamma} + {1 \over 2k_0 r} + i} . \label{eq17} \ee The average, denoted by $\overline{\cdot \cdot  \cdot}$, is  taken over distances $r$ up to a maximal value $r_m \ll k_0
^{-1}$ which accounts for all possible mechanisms that may break  those pairs.  In the considered limit,   the density of the gas can be assimilated to that of the pairs. The imaginary part of $\overline{\Sigma}_{e}
^{(1)}$ defines the elastic mean free path $l_e$ 
by $ {k_0 / l_e} = - \mbox{Im} \overline{\Sigma}_{e} ^{(1)} $, namely
 \be
 {1 \over l_e} = {3 \pi n_i \over k_0 ^2}{1 \over  k_0 r_m}  \int_0^{2k_0 r_m} {dx \over 1 + \left(  { \delta \over \Gamma} + {1 \over x} \right)^2} . 
 \label{eq19}
 \ee
It is interesting to compare $l_e$ to the mean free path $
l_0 = { k_0 ^2 \over 6 \pi n_i}  
\left(1 +  \left( 2  \delta / \Gamma \right)^2 \right) 
$ that corresponds to near resonant elastic scattering of a photon by independent atoms.  At resonance $(\delta =0)$, we have ${ l_0 \over l_e} = {4 \over 3} (k_0 r_m )^2 \ll 1$. Away from resonance, the elastic mean free path $l_e$ becomes smaller than $l_0$ and for blue detuning it is reduced  in a ratio roughly given by $1/ (k_0 r_m)^2$. 

Another  important physical quantity is the group velocity $v_g$ given in terms of the refraction index $\eta$ by $c / v_g = \eta + \omega {d\eta \over d\omega}$. Since $\overline{\Sigma}_{e}
^{(1)}$ is proportional  to the polarizability, 
 the refraction index depends on its real part, namely $
\eta = ( 1 - (c / \omega)^2 \mbox{Re} \overline{\Sigma}_{e}
^{(1)} )^{1/2} 
$. From (\ref{eq17}), we notice that $\eta \simeq 1$ for all values of the detuning $ \delta / \Gamma$ and in a large range of densities  $n_i$ 
so that  
\be
{c \over v_g} = 1- {n_i \over n_c} {1 \over 2 k_0 r_m} f(k_0 r_m, { \delta \over \Gamma})
\label{eq20}
\ee
where we have defined $n_c = {k_0 ^3 \over 6 \pi} {\Gamma \over  \omega_0}$ and the function 
\be
f(k_0 r_m, \Delta) = \int_{0}^{2 k_0 r_m} dx {1 - (\Delta + {1 \over x})^2 \over \left(1 + (\Delta + {1 \over x})^2\right)^2} . 
\label{eq21}
\ee
This expression of $v_g$ diverges at a large and negative value of the detuning ${ \delta \over \Gamma} \simeq -{1 \over  2 k_0 r_m}$ and beyond it takes both positive and negative values. Otherwise it is well behaved, meaning that it remains finite and positive for all values of the density $n_i$. At resonance, the group velocity is \be
{c \over v_g} = 1 + {4 \pi n_i \over k_0 ^3}{\omega_0 \over \Gamma} (k_0 r_m)^2 . 
\label{eq22}
\ee

The present expression of $v_g$ differs substantially from the one obtained for light interaction with independent two-level atoms. There, for densities $n_i > n_c$ where $n_c$ defined above is usually overwhelmingly small,  the group velocity is known to diverge at  two symmetric values of the detuning of order unity and takes negative values in between ({\it i.e.}, also at resonance). For instance, in a gas of $\mbox{Rb}^{85}$ atoms, where $n_i= 6 \times 10^{10} cm^{-3}$, $\lambda_0 = 780 nm$ and ${\Gamma \over 2 \pi } = 5.9 \mbox{MHz}$, we have ${n_i \over n_c} \simeq 10^5$. The validity of the concept of group velocity in such systems has thus been often questioned \cite{brillouin} and an energy velocity has been defined which describes  energy transport through a diffusive medium \cite{bart3}.  

Transport of photons through a diffusing gas is characterized by the diffusion coefficient $D = {1 \over 3} v_g l_e$ that combines the elastic mean free path and the group velocity \cite{am,ishimaru}, both derived from the complex valued self-energy (\ref{eq17}). The diffusion coefficient $D$ is of great importance since it enters in expressions of all measured physical quantities such as reflection and transmission coefficients, angular correlations of speckle patterns, time correlation functions of the intensity (diffusing wave spectroscopy),  {\it \mbox{etc}}. \cite{am}. Moreover, the critical behavior of transport close to Anderson localization transition at strong disorder is also obtained from the scaling form of $D$. Its expression, deduced from (\ref{eq19}) and (\ref{eq20}), depends on the range $r_m$ and on the 
 detuning $ \delta / \Gamma$.  Since the group velocity and the elastic mean free path are significantly modified for  the case of superradiant pairs, we thus expect the diffusion coefficient to be different from its value obtained for  independent atoms. We also define the transport time by $\tau_{tr} (\delta ) = l_e / v_g = {3 D \over v_g ^2} $. At resonance, it  can be rewritten with the help of (\ref{eq22}) as $ \tau_{tr} (0) = {1 \over 2 \Gamma}$ which is consistent with our considering  of superradiant pairs. We would like nevertheless to call attention to the fact that, away from resonance,  $\tau_{tr} (\delta)$ depends on frequency. 

We now compare our results to recent measurements of the diffusion coefficient $D$ and of the group velocity $v_g$ obtained for  multiple scattering of light at resonance, in a cold atomic gas of $\mbox{Rb}^{85}$ \cite{vaujour}. Since the range $r_m$ cannot be directly determined, we first use Eqs. (\ref{eq19}) and (\ref{eq20}) to obtain an expression independent of $k_0 r_m$ given by the ratio ${ (v_g / c)^2 \over 3 D} = {8 \pi n_c \over k_0 ^2 c} = 2 \Gamma / c^2$. For  $\mbox{Rb}^{85}$ atoms, this ratio equals $ 8.2 \times 10^{-10} \mbox{s/m}^2$, which is in good  agreement with the value $4.8 \times 10^{-10} \mbox{s/m}^2$ obtained from measurements of $D$ and $v_g$.  Finally, from the previous numerical expression we deduce for the maximal range of interaction $r_m$  the value $k_0 r_m \simeq 0.51$ also consistent with our assumption of superradiant states. Therefore, multiple scattering of photons by superradiant pairs provides a relevant mechanism that needs to be considered, in addition to others {\it e.g.} scattering by independent atoms,  for description of multiple scattering properties of dilute cold  atomic gases.

 

\medskip

We have considered multiple scattering of a photon on pairs of atoms that are in a superradiant state. On average over disorder configurations, an attractive interaction potential builds up between close enough atoms that decays like $1/r$. The contribution of superradiant pairs, resulting from this potential,  to  scattering properties is  significantly different  from that of independent atoms.  It leads  to a well defined but much smaller group velocity as compared to $c$ and correlatively to a smaller diffusion coefficient. For densities considered in recent experiments on cold $\mbox{Rb}^{85}$, the quantity $k_0 l_e$ that describes eventually the closeness to a localization transition, is reduced at moderate detunings, by one order of magnitude. This effect is expected to be even stronger for larger densities which could then be close to the localization edge. This research is supported in part by the Israel Academy of
Sciences and by the Fund for Promotion of Research at the
Technion.


\begin{thebibliography}{99}
\bibitem{rk} It can be shown using (\ref{eq3}), that the contribution of mixed terms like $ \langle \phi^{\pm}| G | \phi^{\mp}
\rangle$ vanishes.
\bibitem{rk2} This rewriting of the amplitude $T$ is defined up to a proportionality constant which  accounts for the quantum nature of the problem and the photon polarization.
\end{thebibliography}
    \end{document}